\documentclass[lettersize,journal]{IEEEtran}
\usepackage{amsmath,amsfonts}
\usepackage{array}
\usepackage{textcomp}
\usepackage{stfloats}
\usepackage{url}
\usepackage{verbatim}
\usepackage{graphicx}
\usepackage{cite}
\usepackage{xcolor}
\usepackage{algorithm2e}
\SetKwRepeat{Do}{do}{while}%
\RestyleAlgo{ruled}
\usepackage{caption}
\usepackage{subcaption}

\SetCommentSty{mycommfont}

\DeclareMathOperator*{\argmax}{arg\,max}

\IEEEoverridecommandlockouts
\usepackage{tikz}
\usepackage{textcomp}
\usepackage[hidelinks]{hyperref}
\usepackage{lipsum}
\newcommand\copyrighttext{%
  \footnotesize \textcopyright 2025 IEEE. Personal use of this material is permitted.  Permission from IEEE must be obtained for all other uses, in any current or future  media, including reprinting/republishing this material for advertising or promotional  purposes, creating new collective works, for resale or redistribution to servers or  lists, or reuse of any copyrighted component of this work in other works.
  DOI: \href{<http://tex.stackexchange.com>}{10.1109/LPT.2025.3566979}
  }
\newcommand\copyrightnotice{%
\begin{tikzpicture}[remember picture,overlay]
\node[anchor=south,yshift=10pt] at (current page.south) {\fbox{\parbox{\dimexpr\textwidth-\fboxsep-\fboxrule\relax}{\copyrighttext}}};
\end{tikzpicture}%
}

\begin{document}

\title{ORIS allocation to minimize the outage probability\\ in a multi-user VLC scenario}

\author{Borja Genoves Guzman,~\IEEEmembership{Senior Member,~IEEE,}
Maïté Brandt-Pearce,~\IEEEmembership{Fellow,~IEEE}
\thanks{Borja Genoves Guzman has received funding from the European Union under the Marie Skłodowska-Curie grant agreement No 101061853. This work has been partially funded by project PID2023-147305OB-C31 (SOFIA-AIR) by MCIN/ AEI/10.13039/501100011033/ERDF, UE, and by project TUCAN6-CM (TEC-2024/COM-460), funded by CM (ORDEN 5696/2024).}
\thanks{Borja Genoves Guzman is with Universidad Carlos III de Madrid, Depto. Teoría de la Señal y Comunicaciones, Leganés (Madrid), 28911 Spain. Maïté Brandt-Pearce is with University of Virginia, Department Electrical and Computer Engineering, Charlottesville, VA 22904 USA. \\E-mails: bgenoves@ing.uc3m.es, mb-p@virginia.edu}
}



\maketitle

\copyrightnotice 
\vspace{-\baselineskip}

\begin{abstract}
Visible Light Communication (VLC) is a promising solution to address the growing demand for wireless data, leveraging the widespread use of light-emitting diodes (LEDs) as transmitters. However, its deployment is challenged by link blockages that cause connectivity outages. Optical reconfigurable intelligent surfaces (ORISs) have recently emerged as a solution to mitigate these disruptions. This work considers a multi-user VLC system and investigates the optimal association of ORISs to LEDs and users to minimize the outage probability while limiting the number of ORISs used. Numerical results from our proposed optimization algorithm demonstrate that using ORISs can reduce the outage probability by up to 85\% compared to a no-ORIS scenario.
\end{abstract}

\begin{IEEEkeywords}
Line-of-sight (LoS) link blockage, mirrors, multi-user, outage probability, optical reconfigurable intelligent surfaces (ORISs), resource allocation, visible light communication (VLC).
\end{IEEEkeywords}

\section{Introduction}
\IEEEPARstart{V}{isible} light communication (VLC) has emerged as a promising technology to complement traditional radio frequency (RF) systems, addressing the ever-increasing demand for wireless data services.  While the available RF spectrum has become scarce and fragmented, VLC leverages the visible light spectrum, effectively reusing existing lighting infrastructure. However, VLC is susceptible to significant path loss when obstructions arise between the transmitter and receiver.  This issue has been partially mitigated by cooperative techniques~\cite{LightsAndShadows} and relaying schemes~\cite{ReflectionBasedRelayVLC}. In this paper, we exploit reflective surfaces and optimize resource allocation to minimize the outage probability in a multi-user VLC environment.

Optical reconfigurable intelligent surfaces (ORISs) have recently been proposed to provide secondary paths between VLC transmitters and receivers. The flexibility of ORIS elements allows for manipulation of the VLC channel, effectively shielding the link from obstructions.  Two primary types of ORIS elements exist: mirrors and metasurfaces~\cite{MirrorVSMetasurface}.  Due to their superior performance, maturity, and market availability, we focus on mirrors as the ORIS element in this work.  The use of mirrors as ORIS has been shown to improve data rates~\cite{SumRateMirrors}, spectral efficiency~\cite{JointResourceMirrors}, secrecy rates~\cite{ORISSecrecy}, and illumination uniformity~\cite{MirrorVLC}.  However, outage probability, a critical metric significantly impacted by line-of-sight (LoS) link blockage in VLC, has received limited attention~\cite{Globecom2023}\cite{guzman2024resource}.

Unlike previous works that consider outage probability in single-user scenarios as the target for improvement~\cite{Globecom2023}\cite{guzman2024resource}\cite{guzman2024usingcurvedmirrorsdecrease}, this study proposes an ORIS resource allocation algorithm to minimize outage probability in a realistic multi-user scenario.  We formulate an iterative optimization problem that allocates resources only to users likely to be accommodated by the system, thereby avoiding resource wastage. 


\begin{figure}[t]
     \centering
     \begin{subfigure}[t]{0.37\columnwidth}
         \centering
         \includegraphics[width=\textwidth]{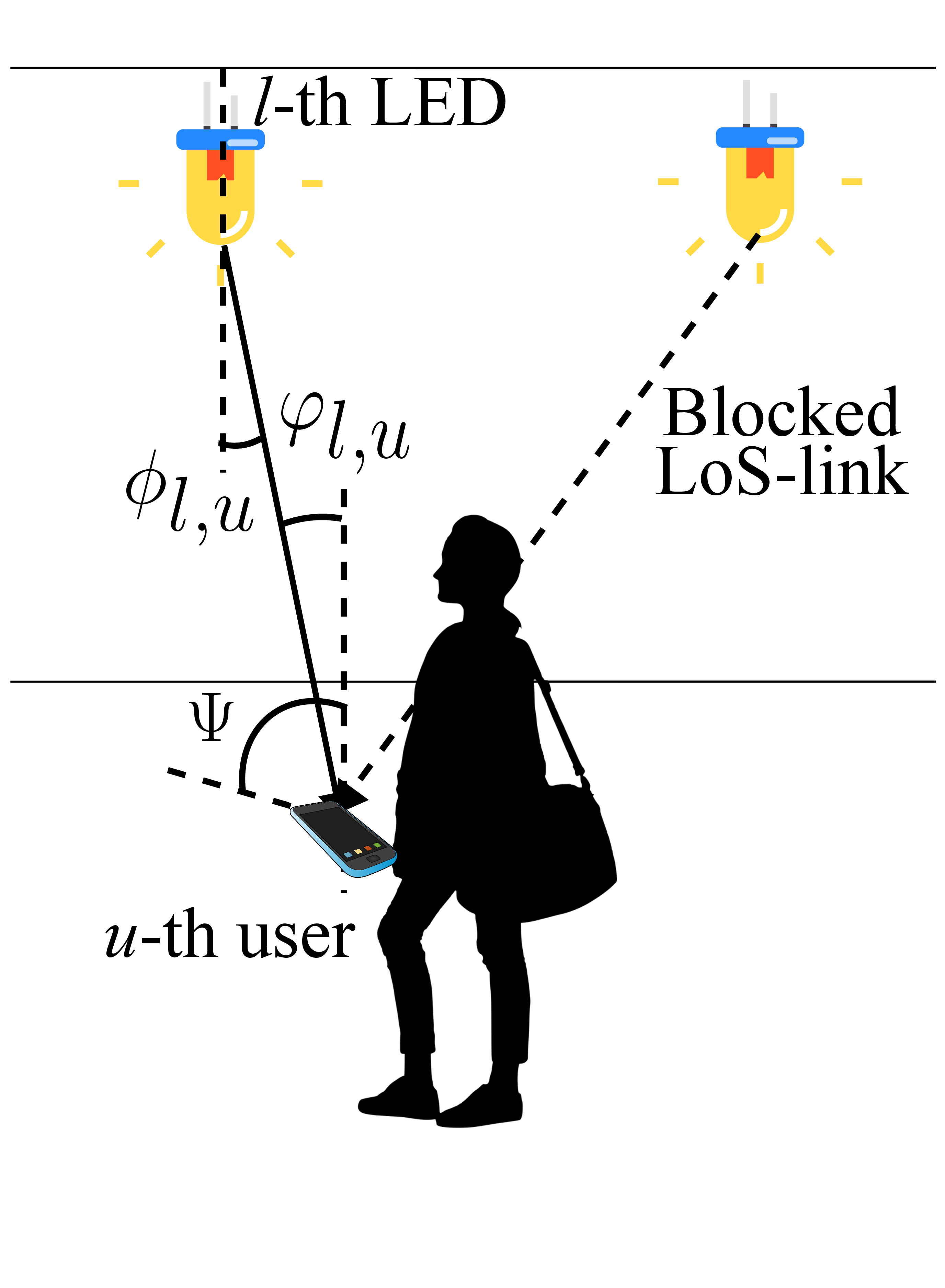}
         \caption{LoS}
         \label{fig:ScenarioLoS}
     \end{subfigure}
     \hfill
     \begin{subfigure}[t]{0.57\columnwidth}
         \centering
         \includegraphics[width=\textwidth]{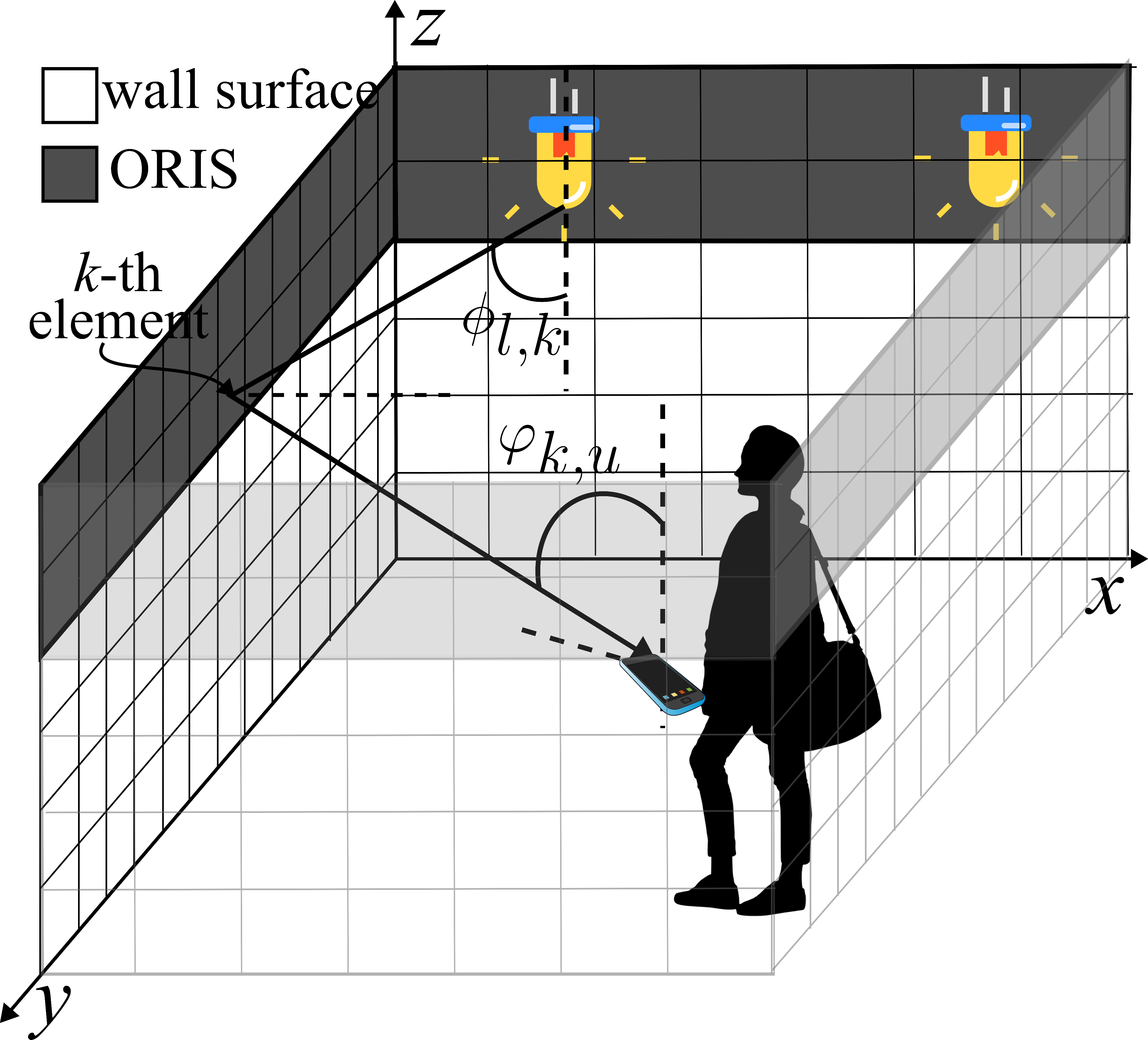}
         \caption{NLoS}
         \label{fig:ScenarioNLoS}
     \end{subfigure}
        \caption{3D illustration of LoS and NLoS propagation. Light sources are located on a horizontal plane on the ceiling. ORIS (mirrors) are placed forming a crown molding in the room.}        
        \label{fig:SystemModel}
        \vspace{-0.7cm}
\end{figure}

\section{System model}
\label{Sec:SystModel}

The considered indoor VLC scenario has $L$ ceiling-mounted and downward-facing LEDs, denoted by $l=\{0,\cdots,L{-}1\}$, and $U$ users distributed in the room, denoted by $u=\{0,\cdots,U{-}1\}$. Each user has an associated photodetector (PD) appointing straight upwards. Due to their small dimensions, LEDs and PDs are modeled as point sources and detectors, respectively. The communication performance relies on LoS and non-line-of-sight (NLoS) links defined as follows.

The LoS channel gain from LED $l$ to user $u$ follows a Lambertian emission model as~\cite{VLCLoSChannel}
\begin{equation}
\label{eq:ChannelModelLoS}
H_{l,u}^{{\rm{LoS}}} {=}\hspace{-1mm}
\begin {cases}
I_{l,u}\frac{{\left( {m + 1} \right) \cdot {A_{{\rm{PD}}}}}}{{2\pi d_{l,u}^2}}{{\cos }^m}\left( {{\phi _{l,u}}} \right)\cos \left( {{\varphi _{l,u}}} \right) & 0 \le {\varphi _{l,u}} \le {\Psi}\\ 
0 & \rm{otherwise},
\end{cases}
\end{equation}
where $m=-1/\log_2\left(\cos\left(\phi_{1/2}\right)\right)$ is the Lambertian index of the LED that models the radiation pattern defined by its half-power semi-angle $\phi_{1/2}$. The parameter $A_{{\rm PD}}$ stands for the active PD area, $d_{l,u}$ is the Euclidean distance between LED $l$ and user $u$, and $\phi_{l,u}$ and $\varphi_{l,u}$ are the irradiance and incidence angles, respectively, as represented in Fig.\,\ref{fig:ScenarioLoS}. The field-of-view (FoV) semi-angle of the PD is denoted by $\Psi$. The variable ${I}_{l,u}$ is Boolean and takes a value of 0 if the LoS link between LED $l$ and user $u$ is obstructed and 1 otherwise.

The NLoS channel considered can be composed of ORIS and wall reflections. In this paper, we consider that each wall is divided into a grid of $K$ ORIS elements, denoted by $k=\{0,\cdots,K{-}1\}$, and $W$ wall elements, denoted by $w=\{0,\cdots,W{-}1\}$, with a total of $K + W$ elements per wall. According to the results obtained in~\cite{guzman2024resource}, the optimal ORIS placement is the upper part of each wall. Therefore, in this paper we consider a ``crown molding'' design where each wall is provided with ORIS elements in the upper parts, as shown in Fig.\,\ref{fig:ScenarioNLoS}. Each ORIS element is installed in a micro-electromechanical structure that allows mirrors to be oriented optimally to reach the receiver. We assume the blockage between reflectors is negligible. The NLoS channel gain produced by an ORIS element $k$ directing the signal from LED $l$ to user $u$ is:
\vspace{-1mm}
\begin{equation}
\label{eq:ChannelModelNLoSRIS}
H_{l,k,u}^{{\rm{ORIS}}} {=}
\begin {cases}
\hspace{-1mm}{{\hat{r}} {\cdot} {\frac{{\left( {m + 1} \right) \cdot {A_{{\rm{PD}}}}}}{{2\pi {{\left( {{d_{l,k}} + {d_{k,u}}} \right)}^2}}}{{\cos }^m}{\left( {{\phi _{l,k}}} \right)}{\cos} \left( {{\varphi _{k,u}}} \right)}} & { 0 {\le} {\varphi _{k,u}} {\le} {\Psi}} \\
\hspace{-1mm}0 & \rm{otherwise}, \\
\end{cases}
\end{equation}
where $d_{l,k}$ and $d_{k,u}$ are the Euclidean distances from LED $l$ and ORIS element $k$, and from  ORIS element $k$ to  user $u$, respectively. Variable $\phi _{l,k}$ is the irradiance angle from LED $l$ to ORIS element $k$, and $\varphi _{k,u}$ is the incidence angle from ORIS element $k$ to user $u$, as represented in Fig.\,\ref{fig:ScenarioNLoS}. The reflection coefficient of an ORIS element is denoted by $\hat{r}$.

The NLoS channel gain produced by a wall grid element $w$ coming from LED $l$ to user $u$ is given by:
\vspace{-1mm}
\begin{equation}
\label{eq:ChannelModelNLoSdiff}
\hspace{-13mm} \begin{aligned}
 & H_{l,w,u}^{{\rm{wall}}} {=} \\
 &  \begin {cases}
{{\tilde{r}} {\cdot} \frac{{\left( {m + 1} \right)  {A_{{\rm{PD}}}}}}{{2{\pi}d_{l,w}^2d_{w,u}^2}}\hspace{-1mm}{A_k}{{\cos }^m}{\left( {{\phi _{l,w}}} \right)}{\cos \left( {{\varphi _{l,w}}} \right)}} { {\cos {\left( {{\phi _{w,u}}} \right)}}{\cos {\left( {{\varphi _{w,u}}} \right)}}} \\& {\hspace{-3cm} 0 {\le} {\varphi _{w,u}} {\le} {\Psi}} \\
0 & \hspace{-3cm}\rm{otherwise}, \\ 
\end{cases}
\end{aligned}
\vspace{-1mm}
\end{equation}
where $\tilde{r}$ is the reflection coefficient of a wall surface, $\varphi _{l,w}$ is the incidence angle onto wall element $w$ coming from LED $l$, and $\phi _{w,u}$ is the irradiance angle from element $w$ to user $u$.

Due to Snell's law, each ORIS element can redirect the light coming from one LED $l$ to a single user $u$. Therefore, we must propose a resource allocation strategy to associate each ORIS element $k$ with an LED $l$ and user $u$. This is modeled with a Boolean variable defined as
\begin{equation}
\beta _{l,k,u} = \begin {cases}
1 & \text{if $k$ is associated with LED $l$ and user $u$} \\
0 & \text{otherwise}. \\
\end{cases}
\end{equation}
Thus, the NLoS channel gain from LED $l$ to user $u$ is
\vspace{-1mm}
\begin{IEEEeqnarray}{rCl}
\hspace{-1mm} H_{l,u}^{{\rm{NLoS}}}\left( {{\beta _{l,k,u}}} \right) & {=} & \hspace{-1mm} \sum_{w=0}^{W{-}1} \hspace{-1mm} I_{l,w,u} H_{l,w,u}^{{\rm{wall}}}  {+} \hspace{-1mm} \sum_{k=0}^{K{-}1} \hspace{-1mm} I_{l,k,u} H_{l,k,u}^{{\rm{ORIS}}} {\beta _{l,k,u}},
\label{eq:ChannelModelNLoS}
\end{IEEEeqnarray}
where $I_{l,w,u}$ and $I_{l,k,u}$ are Boolean variables that take a value of 0 when the NLoS link from LED $l$ passing through reflecting element $w$ or $k$ to user $u$ is blocked, respectively, and 1 otherwise. Blockage may be generated by the user's own body and by other users. Therefore, the LED-ORIS-user association denoted by $\beta_{l,k,u}$ is key to the system performance.

The overall VLC channel gain from LED $l$ to the user $u$ can be formulated as
\begin{IEEEeqnarray}{rCl}
H_{l,u}\left( {{\beta _{l,k,u}}} \right) & {=} & H_{l,u}^{{\rm{LoS}}} + H_{l,u}^{{\rm{NLoS}}}\left( {{\beta _{l,k,u}}} \right).
\label{eq:ChannelModelTotal}
\end{IEEEeqnarray}

To eliminate the possibility of interference, we assume a multicarrier modulation scheme where subcarriers are uniquely assigned to users, with the number of subcarriers at least as large as the number of users. Let us further assume a single access point with distributed LEDs that have an optical power per LED and subcarrier of $P_{\rm sc}$. Therefore, the signal-to-noise-power ratio (SNR) of user $u$ can be formulated as
\begin{IEEEeqnarray}{l}
    {\gamma}_u({\boldsymbol{\beta}})  = 
    \frac{{{{\left( {{{\rho\cdot P_{\rm sc}}\sum\limits_l H_{l,u}\left( \beta _{l,k,u}\right)} } \right)}^2}}}{{{N_0}B}}, 
\label{eq:SNR}
\end{IEEEeqnarray}
where $\boldsymbol{\beta}$ is an $L \times 4K \times U$ matrix containing all $\beta_{l,k,u},\,\forall l,k,u$ values. The parameter $\rho$ is the PD responsivity, $N_0$ is the power spectral density of the additive white Gaussian noise (AWGN) at the receiver mainly produced by shot and thermal noise~\cite{VLCLoSModel}, and $B$ is the bandwidth. 

The outage probability is defined as the probability that a user's subcarrier has an SNR lower than the required SNR, called the SNR threshold ($\gamma_{\rm th}$), and can be written as
\begin{IEEEeqnarray}{rCl}
P_{{\rm out}}({\boldsymbol{\beta}},\gamma_{{\rm th}}) & = &  \Pr\{\gamma_u({\boldsymbol{\beta}})<\gamma_{{\rm th}}\}.
\label{eq:OutProb}
\end{IEEEeqnarray}

\section{Optimization problem formulation}
\label{Sec:OptimizationProblem}

To minimize the outage probability in an ORIS-assisted VLC scenario, we propose a resource allocation algorithm that maximizes the minimum SNR among all users for which the minimum SNR exceeds $\gamma_{\rm th}$, i.e., we do not assign resources to users that remain in outage despite the assistance of the ORIS. Thus, this algorithm is as fair as possible as it minimizes the differences in the communication performance of all users that are not in outage.\footnote{If VLC supplements an RF-based system, the RF resources can be dedicated to users that experience VLC-outage. If not, then the VLC system may dedicate some portion of time to just accommodate those hard-to-reach users in a time-division manner. If the users in outage cannot be served by the VLC system no matter what, they will remain in outage until they move and the algorithm reoptimizes.}

Let us assume we have a DCO-OFDM modulation scheme composed of $N$ subcarriers. The total optical power per LED is $P_{\rm tot} = P_{\rm sc} \times \sqrt{N-2}$ because the first and N/2-th subcarriers do not transmit data.  
We propose obtaining the optimal LED-ORIS-user association ($\boldsymbol{\beta^*}$) by solving the following optimization problem:
\vspace{-2mm}
\begin{IEEEeqnarray}{l}
\label{MINOUT_MultiUser}
\boldsymbol{\beta^*} = \mathop {\mathop {\argmax } \limits_{{\boldsymbol{\beta}}} \left(\min\limits_u \gamma'_u(\boldsymbol{\beta}) -\epsilon\sum\limits_{l,k,u} {{\beta _{l,k,u}}}\right)}\limits\\
\begin{array}{l}
{\rm{subject\,to}}\\
{\rm{C1:  }}\,\sum\limits_l\sum\limits_u {{\beta _{l,k,u}} \le {1}}, \forall \, k \\
{\rm{C2:  }}\,{\beta _{l,k,u}} \in \{ 0,1\} ,{\rm{  }}\, \forall \, l,k,u\\
\end{array} \IEEEnonumber
\end{IEEEeqnarray}
For mathematical simplicity, we utilize the optical SNR defined as ${\gamma}'_u({\boldsymbol{\beta}})  = \sqrt{{\gamma}_u({\boldsymbol{\beta}})}$ in our objective function, keeping it as a linear function of the optimization variables.
\begin{algorithm}[t]
\caption{Proposed resource allocation algorithm}\label{alg:ResourceSaving}
\KwData{$\rho, P_{\rm sc}, N_0, B_{\rm sc}, \gamma_{\rm th}', H_{l,u}^{{\rm{LoS}}}, H_{l,k,u}^{{\rm{ORIS}}}, H_{l,w,u}^{{\rm{wall}}}, I_{l,w,u}, \newline I_{l,k,u}, \forall\,l,k,w,u$}
\KwResult{$\boldsymbol{\beta^*}, \gamma_{\rm min}'^*$}
\DontPrintSemicolon
Solve \eqref{MINOUT_MultiUserv2} \tcp*{\textcolor{blue}{Allocate resources to all users}}
\tcp*{\textcolor{blue}{Check if user with lowest SNR is in outage:}}
\While{$\gamma_{\rm min}'^*< \gamma_{\rm th}'$}{
$\mathcal{U}=\mathcal{U}\backslash\Bar{u}:\gamma_{\Bar{u}}'=\gamma_{\rm min}'^*$\tcp*{\textcolor{blue}{Such user is removed from the set of supported users}}
Solve \eqref{MINOUT_MultiUserv2} \tcp*{\textcolor{blue}{Reallocate resources to remaining users}}}
\end{algorithm}
Constraint C1 restricts the allocation of every ORIS element to a single LED and user. The infinitesimal value $\epsilon$ (set to 10$^{-3}$ in the numerical results) forces the algorithm toward a unique optimal solution that minimizes the number of ORISs employed without compromising the optimality of the outage probability. 

We revise the objective function by introducing a new decision variable $\gamma_{\rm min}'=\min\limits_u \gamma'_u$, where $u\in\mathcal{U}=\{0,1,...,U-1\}$. In addition, we propose an iterative algorithm to prevent the system from wasting resources on users that are in outage in any case. This solvable optimization algorithm is detailed in Algorithm\,\ref{alg:ResourceSaving}, where \eqref{MINOUT_MultiUserv2} in there is
\vspace{-0.3cm}
\begin{IEEEeqnarray}{l}
\label{MINOUT_MultiUserv2}
[\boldsymbol{\beta^*}, \gamma_{\rm min}'^*] = \mathop {\mathop {\argmax } \limits_{{\boldsymbol{\beta}}, \gamma_{\rm min}'}  \left(\gamma_{\rm min}' -\epsilon\sum\limits_{l,k,u} {{\beta _{l,k,u}}} \right)}\\
\begin{array}{l}
{\text{subject\,to\,C1,\,C2\,and}}\\
{\rm{C3:  }}\,{\gamma_{\rm min}'\leq \gamma_u'(\boldsymbol{\beta})}, \, \forall \, u\\
\end{array} \IEEEnonumber
\end{IEEEeqnarray}
Note that this Algorithm\,\ref{alg:ResourceSaving} allocates ORIS elements to users optimally in each iteration and repeatedly checks if the user is still in outage. In this case, the user will not be supported and its allocated resources are released to be allocated to the rest of the users in the next iteration. The algorithm stops when resources are allocated among the users that are not in outage, i.e., when those resources are employed usefully. Algorithm\,\ref{alg:ResourceSaving} must run every time the channel conditions change in order to allocate resources optimally.

The problem introduced is a Linear Programming (LP) problem. In practice, the complexity order can be approximated as $O(v^2c)$ through big-$O$ notation, where $v=4\cdot L\cdot K \cdot U$ is the number of variables and $c=4\cdot K+U$ is the number of constraints~\cite{boyd_vandenberghe_2004}. Note that the iteration process of Algorithm\,\ref{alg:ResourceSaving} does not affect the order of complexity. Then, the total complexity of our proposed resource allocation algorithm can be approximated as $O((L\cdot K\cdot U)^2(K+U))$, which can be solved with a solver such as Gurobi and CVX~\cite{cvx}.

\vspace{-2mm}
\section{Simulation results}
\label{Sec:SimulationResults}
\vspace{-1mm}
This section presents a comprehensive analysis of the simulation results for the proposed VLC system, which incorporates an ORIS  occupying the upper one-third of each wall. To ensure general applicability, we consider a residential space with dimensions of 4$\times$4$\times$3 meters, where a total of $L=4$ LEDs are arranged in a symmetric 2-by-2 grid at coordinates [1,1], [1,3], [3,1], and [3,3] meters. NLoS reflections from all four walls are accounted for, both ORIS and wall elements. Unless otherwise stated, the number of ORIS deployed per wall is $K=30\times 5=150$.

Given the deterministic nature of the VLC channel, we adopt an object-based model to account for link-path blockages~\cite{CylinderBlockage}. This model incorporates both self-blocking and inter-user blocking effects. The user’s body is represented as a cylindrical object with a height of 1.75\,m and a radius of 0.15\,m. The user holds a device equipped with a single PD facing upwards, positioned 0.3\,m away from the body at a height of 1\,m. Additionally, the user's equipment is oriented at a random angle on the horizontal plane with respect to the cylinder that follows a uniform distribution from 0 to $2\pi$~\cite{CylinderBlockage}.

We analyze and compare three algorithms: the one proposed in \eqref{MINOUT_MultiUserv2} with ORIS elements possibly allocated to users even if they are in outage, the one proposed in Algorithm\,\ref{alg:ResourceSaving}, and the case where there are no ORISs allocated to users.

\begin{figure}[t]
\centering
\includegraphics[width=\columnwidth]{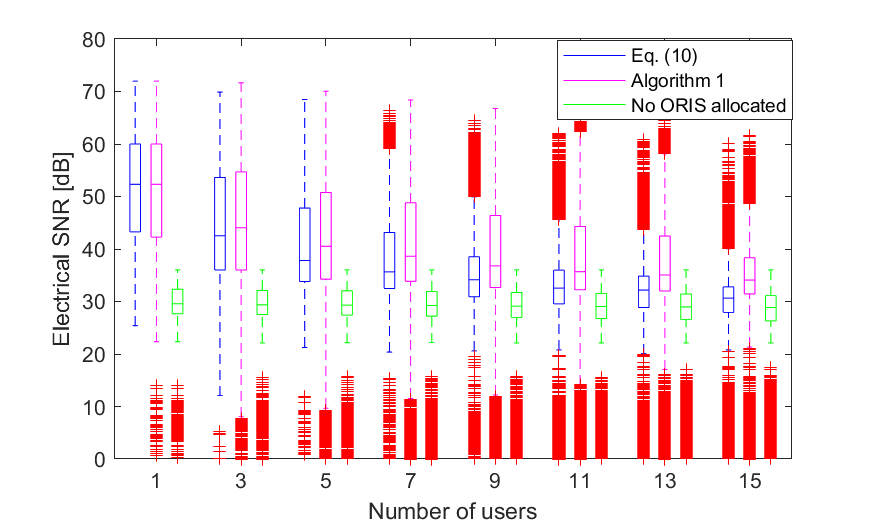}
        \caption{Box and whisker plot of the SNR received for multiple number of users deployed. Parameters: $\tilde{r} = $  0.4, $\hat{r}=$ 0.95, $N=$ 512, $B=$ 20\,MHz, $P_{\rm tot}=$ 10\,W, $N_0= 2.5\cdot$10$^{-20}$\,W/Hz,  $\Psi= 40^\circ$, $A_{\rm PD}=$ 1\,cm$^2$, $\phi_{1/2}=80^\circ$, $\rho=0.4$\,A/W.}
\label{fig:fig1}
\vspace{-0.5cm}
\end{figure}

\begin{figure}[t]
\centering
\includegraphics[width=\columnwidth]{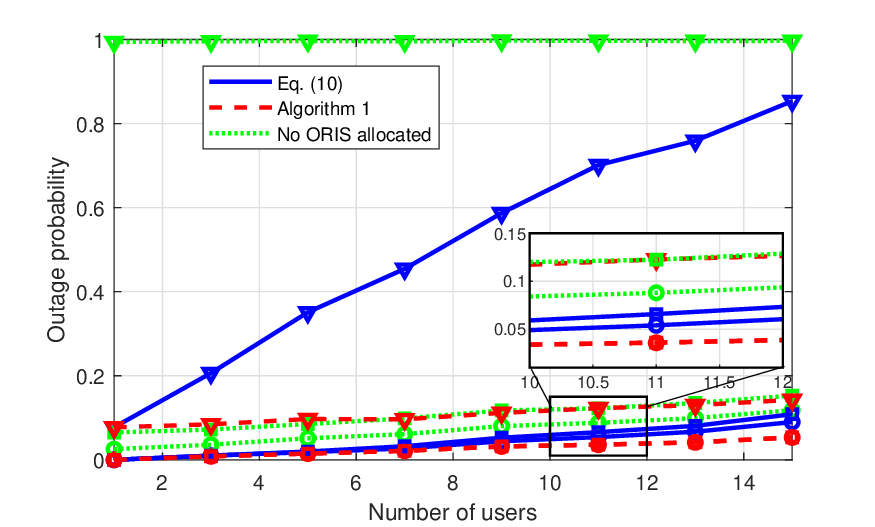}
        \caption{Outage probability for different number of users when $\gamma_{\rm th}=5$\,dB, $\gamma_{\rm th}=20$\,dB and $\gamma_{\rm th}=35$\,dB, which are represented by symbols O, $\square$ and $\nabla$, respectively.}
\label{fig:fig2}
\vspace{-0.6cm}
\end{figure}

We first evaluate the SNR received by a user when a different number of users are deployed, shown in Fig.\,\ref{fig:fig1}. Note that all results presented in this paper use the parameters defined in the caption of Fig.\,\ref{fig:fig1}. As expected, the larger the number of users, the lower the SNR in general, as more blockages may occur and fewer ORIS elements per user are available. However, our proposed Algorithm\,\ref{alg:ResourceSaving} is less degraded compared to the other algorithms because resources are more efficiently allocated, supporting users only if they will not be in outage. To obtain these results we have taken random values of $\gamma_{\rm th}$ uniformly distributed in the range $[0, 50]$\,dB.

\begin{figure}[t]
\centering
\includegraphics[width=\columnwidth]{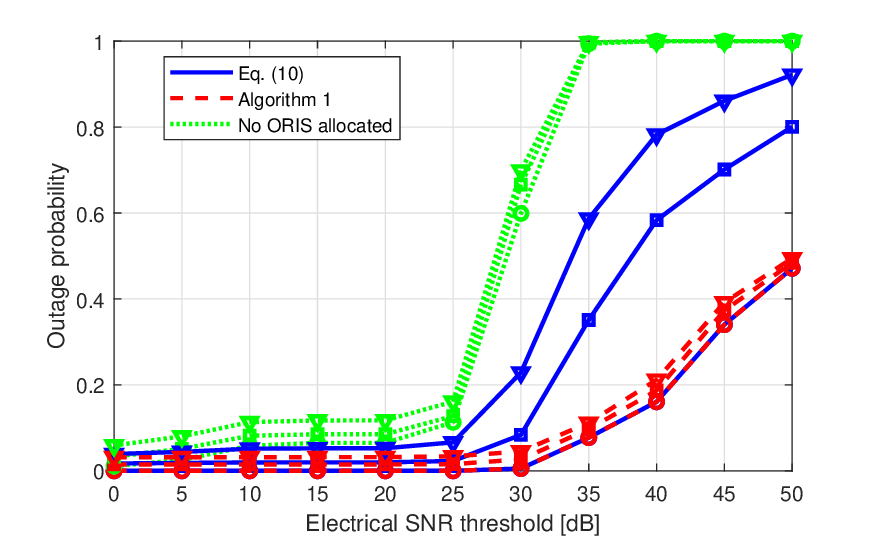}
        \caption{Outage probability as a function of the SNR threshold $\gamma_{\rm th}$, when the number of users deployed is 1, 5 or 9, which are represented by symbols O, $\square$ and $\nabla$, respectively.}
\label{fig:fig3}
\vspace{-0.5cm}
\end{figure}

In Fig.\,\ref{fig:fig2}, we analyze the outage probability for $\gamma_{\rm th}=5$\,dB, $\gamma_{\rm th}=20$\,dB and $\gamma_{\rm th}=35$\,dB, when multiple users are deployed. Again, the larger the number of users, the larger the outage probability as the number of blockages increases and the number of available ORIS elements per user decreases. However, Algorithm\,\ref{alg:ResourceSaving} obtains an outage probability decrease of 85\% and 82\% with respect to no ORIS allocated and the algorithm in \eqref{MINOUT_MultiUserv2}, respectively, when $\gamma_{\rm th}=35$\,dB. When analyzing all possible $\gamma_{\rm th}$ values, shown in Fig.\,\ref{fig:fig3}, our Algorithm\,\ref{alg:ResourceSaving} always provides an outage probability lower than the one given by the other two algorithms. Besides, it is very robust to blockages, as the results obtained are very similar regardless of the number of users deployed.

\begin{figure}[t]
\centering
\includegraphics[width=\columnwidth]{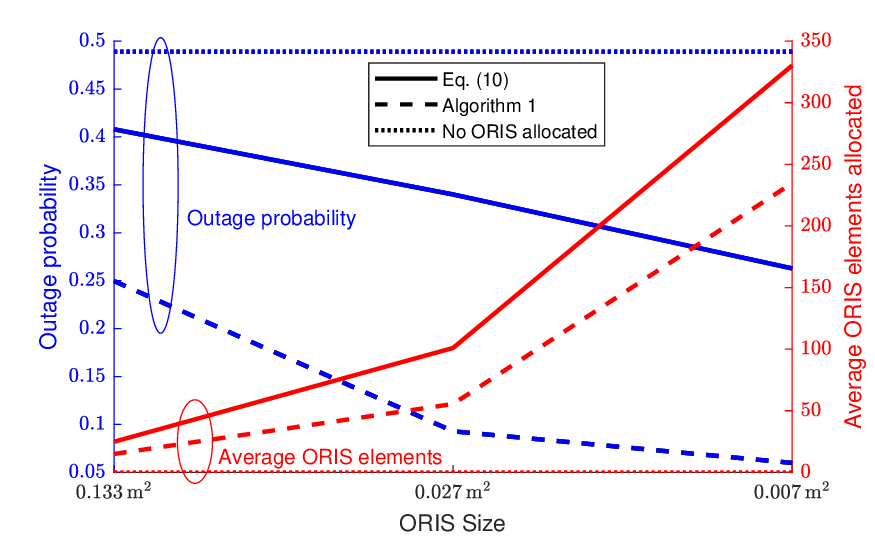}
        \caption{Outage probability and average number of ORIS elements allocated for different sizes of ORIS elements.}
\label{fig:fig4}
\vspace{-0.5cm}
\end{figure}

Finally, in Fig.\,\ref{fig:fig4} we analyze the effect of the size of the ORIS elements on the outage probability and on the average number of ORIS elements allocated. We provide results averaged over all possible numbers of users deployed (1 - 15)  and all $\gamma_{\rm th}$ values  (0 - 50 dB) evaluated in the results above. We analyze three ORIS sizes, 0.133\,m$^2$, 0.027\,m$^2$ and 0.007\,m$^2$, which corresponds to deploying $K$=15$\times$2=30, $K$=30$\times$5=150 and $K$=90$\times$20=1800 ORIS elements per wall, respectively. Note that the smaller the ORIS element (i.e., the larger the number of ORIS elements), the lower the outage probability, as we have more NLoS links to support users. However, the number of ORIS elements allocated to users increases, which leads to a higher system complexity. Despite that, Algorithm\,\ref{alg:ResourceSaving} offers the lowest outage probability and the lowest number of ORIS elements allocated (except for the algorithm with no ORIS allocated), which shows a larger efficiency in resource allocation compared to the other algorithms.


\vspace{-3mm}
\section{Conclusion}
\label{Sec:Conclusion}
\vspace{-1mm}
This study presents an optimization algorithm designed to minimize the outage probability for multiple VLC users within a shared indoor environment by strategically allocating ORIS elements.  The algorithm optimizes both the number of employed ORIS elements and their placement along the upper wall regions.  Numerical results demonstrate a significant reduction in outage probability compared to scenarios without ORIS deployment and those employing less efficient resource allocation strategies.  Furthermore, our algorithm achieves superior SNR performance while simultaneously minimizing the number of required ORIS elements. The robustness of the proposed algorithm against LoS link blockage in multi-user scenarios has also been validated.

\vspace{-2mm}
\section*{Acknowledgments}
\vspace{-1mm}
The authors would like to acknowledge the help of Dr. Victor P. Gil Jimenez, Dr. Maximo Morales Cespedes, and Dr. Ana Garcia Armada for their guidance in previous works where we introduced the problem for minimizing the outage probability in a single-user scenario.

\vspace{-2mm}
\bibliography{./references}

\begin{thebibliography}{10}
\providecommand{\url}[1]{#1}
\csname url@samestyle\endcsname
\providecommand{\newblock}{\relax}
\providecommand{\bibinfo}[2]{#2}
\providecommand{\BIBentrySTDinterwordspacing}{\spaceskip=0pt\relax}
\providecommand{\BIBentryALTinterwordstretchfactor}{4}
\providecommand{\BIBentryALTinterwordspacing}{\spaceskip=\fontdimen2\font plus
\BIBentryALTinterwordstretchfactor\fontdimen3\font minus
  \fontdimen4\font\relax}
\providecommand{\BIBforeignlanguage}[2]{{%
\expandafter\ifx\csname l@#1\endcsname\relax
\typeout{** WARNING: IEEEtran.bst: No hyphenation pattern has been}%
\typeout{** loaded for the language `#1'. Using the pattern for}%
\typeout{** the default language instead.}%
\else
\language=\csname l@#1\endcsname
\fi
#2}}
\providecommand{\BIBdecl}{\relax}
\BIBdecl

\bibitem{LightsAndShadows}
M.~M. Céspedes, B.~G. Guzmán, and V.~P.~G. Jiménez, ``{Lights and shadows: A
  comprehensive survey on cooperative and precoding schemes to overcome LoS
  blockage and interference in indoor VLC},'' \emph{Sensors}, vol.~21, no.~3,
  2021.

\bibitem{ReflectionBasedRelayVLC}
B.~G. Guzmán, C.~Chen, V.~P.~G. Jiménez, H.~Haas, and L.~Hanzo,
  ``{Reflection-based relaying techniques in visible light communications: Will
  it work?}'' \emph{IEEE Access}, vol.~8, pp. 80\,922--80\,935, 2020.

\bibitem{MirrorVSMetasurface}
A.~M. Abdelhady, A.~K.~S. Salem, O.~Amin, B.~Shihada, and M.-S. Alouini,
  ``Visible light communications via intelligent reflecting surfaces:
  Metasurfaces vs mirror arrays,'' \emph{IEEE Open J. Commun. Soc.}, vol.~2,
  pp. 1--20, 2021.

\bibitem{SumRateMirrors}
S.~Sun, F.~Yang, and J.~Song, ``Sum rate maximization for intelligent
  reflecting surface-aided visible light communications,'' \emph{IEEE Commun.
  Lett.}, vol.~25, no.~11, pp. 3619--3623, 2021.

\bibitem{JointResourceMirrors}
S.~Sun, F.~Yang, J.~Song, and Z.~Han, ``Joint resource management for
  intelligent reflecting surface–aided visible light communications,''
  \emph{IEEE Trans. Wireless Commun.}, vol.~21, no.~8, pp. 6508--6522, 2022.

\bibitem{ORISSecrecy}
F.~Wang \emph{et~al.}, ``{Enhancing secrecy of indoor optical RIS aided SSK VLC
  downlink},'' \emph{IEEE Trans. Wireless Commun.}, pp. 1--1, 2025.

\bibitem{MirrorVLC}
S.~Ibne~Mushfique, A.~Alsharoa, and M.~Yuksel, ``{MirrorVLC: Optimal mirror
  placement for multielement VLC networks},'' \emph{IEEE Trans. Wireless
  Commun.}, vol.~21, no.~11, pp. 10\,050--10\,064, 2022.

\bibitem{Globecom2023}
B.~G.~Guzman, M.~M.~Cespedes, V.~P. G.~Jimenez, A.~G.~Armada, and
  M.~Brandt-Pearce, ``Optimal mirror placement to minimize the outage area in
  visible light communication,'' in \emph{Proc. IEEE GLOBECOM}, 2023.

\bibitem{guzman2024resource}
B.~G. Guzman, M.~M. Cespedes, V.~P. Gil~Jimenez, A.~G. Armada, and
  M.~Brandt-Pearce, ``Resource allocation exploiting reflective surfaces to
  minimize the outage probability in {VLC},'' \emph{IEEE Trans. Wireless
  Commun.}, pp. 1--1, 2025.

\bibitem{guzman2024usingcurvedmirrorsdecrease}
B.~G. Guzman, A.~G. Armada, and M.~Brandt-Pearce, ``{On using curved mirrors to
  decrease shadowing in VLC},'' \emph{arXiv:2409.03378}, 2024.

\bibitem{VLCLoSChannel}
J.~Kahn and J.~Barry, ``Wireless infrared communications,'' \emph{Proc. IEEE},
  vol.~85, no.~2, pp. 265--298, 1997.

\bibitem{VLCLoSModel}
T.~Komine and M.~Nakagawa, ``{Fundamental analysis for visible-light
  communication system using LED lights},'' \emph{IEEE Trans. Consum.
  Electron.}, vol.~50, no.~1, pp. 100--107, 2004.

\bibitem{boyd_vandenberghe_2004}
S.~Boyd and L.~Vandenberghe, \emph{Convex Optimization}.\hskip 1em plus 0.5em
  minus 0.4em\relax Cambridge University Press, 2004.

\bibitem{cvx}
M.~Grant and S.~Boyd, ``{CVX}: Matlab software for disciplined convex
  programming, version 2.1,'' \url{http://cvxr.com/cvx}, Mar. 2014.

\bibitem{CylinderBlockage}
S.~Jivkova and M.~Kavehrad, ``Shadowing and blockage in indoor optical wireless
  communications,'' in \emph{Proc. GLOBECOM '03 (IEEE Cat. No.03CH37489)},
  vol.~6, 2003, pp. 3269--3273 vol.6.

\end{thebibliography}
\bibliographystyle{IEEEtran}

\vfill

\end{document}